\documentclass[11pt,twoside]{article}

\usepackage{asp2006}
\usepackage{natbib}
\usepackage{epsf}
\usepackage{lscape}
\usepackage{graphicx}

\begin{document}
\title{The Simulation of Realistic BiSON-like Helioseismic Data}   
\author{Stephen Fletcher and Roger New}   
\affil{Sheffield Hallam University, Sheffield, UK}    
\author{Anne-Marie Broomhall, William Chaplin and Yvonne Elsworth}   
\affil{University of Birmingham, Birmingham, UK}    

\begin{abstract} 
When simulating full-disc helioseismic data, instrumental noise has
traditionally been treated as time-independent. However, in reality,
instrumental noise will often vary to some degree over time due to
line of sight velocity variations and possibly degrading hardware.

Here we present a new technique for simulating Birmingham Solar
Oscillations Network (BiSON) helioseismic data with a more realistic
analogue for instrumental noise. This is achieved by simulating the
potassium solar Fraunhoffer line as observed by the BiSON
instruments. Intensity measurements in the red and blue wing of the
line can then be simulated and appropriate time-dependent
instrumental noise can be added. The simulated time-series can then
be formed in the same way as with real data. Here we present the
simulation method and the first generation of a BiSON-like
instrumental noise time series.
\end{abstract}

\section{Introduction}

Simulating a realistic solar oscillations signal is an important
task within Helioseismology. The use of artificial data, whose
statistical characteristics and input parameters are known, allows
for thorough testing, and may reveal biases in, different analysis
techniques. Most simulations to date have concentrated on accurately
replicating the solar oscillation and background signals, with less
thought put into simulating the instrumental noise sources
associated with the collection of the data. Traditionally,
instrumental noise has simply been treated as a constant
time-independent source. However, in reality instrumental noise
varies over time due to line of sight velocity variations -- at
least for resonant scattering spectrometers (RSS) commonly used to
make full-disc measurements -- and hardware changes. The current
work seeks to introduce realistic simulations of instrumental noise
into standard solar oscillation simulations.

\section{Method}

In order to accurately simulate instrumental noise, we must first
understand how the oscillations are observed. At each of the six
BiSON stations an RSS is used to measure the Doppler shift of the
770-nm solar absorption line. Each RSS contains a cell of Potassium
atoms held at about $100^\textrm{\footnotesize{o}}$C. Detectors are
placed at right-angles to the cell so that only resonantly scattered
photons should be counted. The absorption linewidth of the vapour
cell is much smaller than the width of the Fraunhofer line because
the temperature is much lower and there is no rotational broadening.
Therefore, only light from a narrow band of the solar absorption
line is detected by the RSS. By applying a magnetic field and making
use of the Zeeman effect, the absorption line of the potassium atoms
in the lab is split. Hence, by switching the state of circular
polarisation of the input solar light, it is possible to measure the
light intensity in one wing at a time.

As the solar absorption line undergoes a Doppler shift due to the
orbital motion and spin of the Earth and due to solar oscillations,
so the intensity measurements in the wings change. From these
measurements a ratio, R, is formed to give a near-linear proxy for
the velocity shift of the line:
\begin{equation}
R = \frac{I_b - I_r}{I_b + I_r} =
\frac{I_b^{res}+I_b^{non}+I_b^{elec}+i_b-I_r^{res}-I_r^{non}-I_r^{elec}-i_r}
{I_b^{res}+I_b^{non}+I_b^{elec}+i_b+I_r^{res}+I_r^{non}+I_r^{elec}+i_r}
\label{eq1}
\end{equation}
where $I_b$ and $I_r$ are the intensities in the blue and red wings
of the line respectively. Background offset and instrumental noise
will affect the intensity measurements used to form $R$ so the $I$'s
can be expanded out into $I^{res}$ which is the desired contribution
from resonantly scattered light and $I^{non}$, $I^{elec}$ and $i$
which are background sources due to non-resonantly scattered light,
electronic offsets and noise respectively. In general these are all
functions of time.

To obtain the velocity measurements of the solar oscillations, a
third-order polynomial function of the station-Sun line-of sight
velocity is fitted to $R$. The oscillations signal is then recovered
by subtracting $R$ from the polynomial function and calibrated using
the fitted gradient of $R$ versus station velocity.

The first step in the simulation process was to estimate $I^{res}$
by fitting a Gaussian profile to an observed line obtained using
Doppler velocity observations of the centre of the solar disc. The
observations were taken by the Themis solar telescope located at
Iza$\tilde{\textrm{n}}$a, Tenerife [private communication with
Roasaria Simonello]. The effects of solar rotation, limb darkening
and Doppler imaging where all taken into account in order to
generate a simulated line similar to that seen by the RSS's at each
of the BiSON stations \citep[see][]{Broomhall2007}. The use of this
more realistic line shape is an enhancement on earlier simulation
work reported by \cite{Chaplin2005a}.

\begin{figure}
\centerline{\includegraphics[width=3.4in]{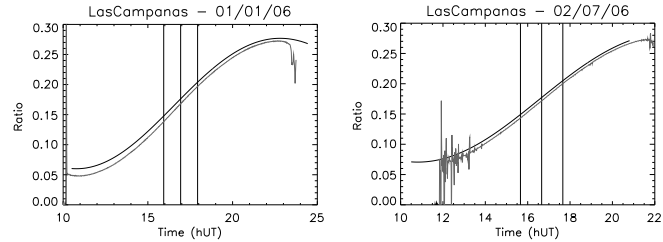}}\caption{Daily
ratio curves for Las Campanas on two different dates. Real data are
given by the grey line and simulated data by the black line.}
\label{RatioPlot}
\end{figure}

The ``operating point" on the simulated line is evaluated from the
Doppler shift due to the changing line of sight velocity of the Sun
to each station (gravitational red shifts and convective blue shift
are included). Artificial velocity oscillations can then be added to
the velocity Doppler shift \citep{Chaplin2008}. Intensity
``measurements" are made for $I_b^{res}$ and $I_r^{res}$ in the same
way as with real data. At this point estimates of the various noise
sources can easily be included. Finally, the ratio can be formed and
analysed in the same way as with real data. A comparison of the
daily ratio curve generated via the simulator compared with that
from real data is shown in Fig.~\ref{RatioPlot}.

\section{Preliminary Results}

\begin{figure}
\centerline{\includegraphics[width=3.4in]{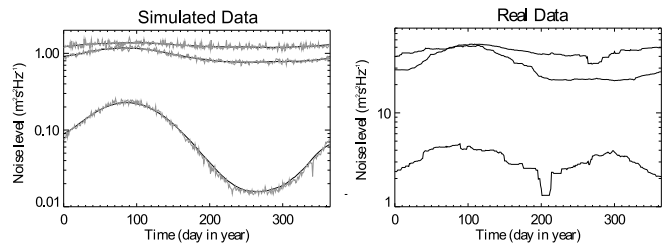}}\caption{Left
panel: Yearly variation in the power spectral density for correlated
noise (bottom), uncorrelated noise (middle) and shot noise (top).
The solid line give the moving mean over 50 days. Right panel: The
moving mean over 50 days of the yearly noise variation in the
instruments at Sutherland (bottom), Carnarvon (middle) and
Iza$\tilde{\textrm{n}}$a (top).} \label{Noise}
\end{figure}

Different noise sources can be categorised by whether the noise they
generate is correlated or uncorrelated in the two wings of the
Fraunhofer line. Noise that varies faster than the switching between
the intensity measurements in the blue and red wings will likely be
uncorrelated, while more slowly fluctuating noise will be
correlated. Using the simulator the effects of these different types
of noise can be tracked. The left panel of Fig.~\ref{Noise} shows
the power spectral density of three different types of noise over
the course of a year. The three cases represent uncorrelated noise
with a constant amplitude, correlated noise with the same constant
amplitude and shot noise (which is equivalent to uncorrelated noise
with an amplitude proportional to $I$). The resulting noise power
clearly varies throughout the year, with the correlated noise
showing the largest variation and the shot noise the smallest.
Although its variation is largest, the correlated noise case
generates somewhat smaller absolute noise levels than the
uncorrelated case.

The simulated noise can be compared with noise in real BiSON data.
The right hand plots in Fig.~\ref{Noise} show the mean power
spectral density of BiSON observations for frequencies from 8.0 to
12.5 mHz (well above the region of the solar oscillations), over a
course of a year. The Carnarvon plot shows relatively high noise and
a large variation similar to the correlated case.
Iza$\tilde{\textrm{n}}$a also has fairly high noise level, but the
variation over the year is fairly small indicating that most
prevalent noise source may be shot noise. Sutherland has a much
lower noise level but, unlike the simulations,  its variation over
the year shows 2 maxima. A possible explanation for this is that the
instrumental noise is actually following a similar trend to the
other plots, but the high frequency ``tail" of the solar background
maybe increasing the observed noise level in the 2nd half of the
year.

\begin{figure}
\centerline{\includegraphics[width=3.2in]{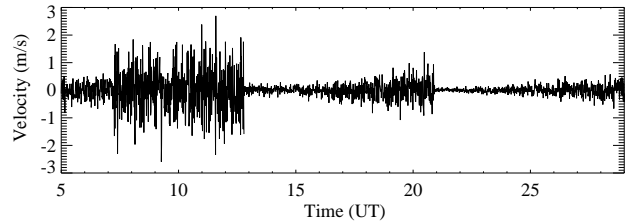}}\caption{Simulated
velocity residuals on July 1st and 2nd 2007 for four different BiSON
stations. Neither solar oscillations nor background were included
allowing for an unobstructed plot of the instrumental noise. The
time series is generated assuming the residuals with lower noise are
chosen when data overlaps.} \label{ResidStitch}
\end{figure}

Instrumental noise also varies on a daily basis, again due to the
change in operating point on the solar line. This is very difficult
to observe with real data due to the presence of the solar
oscillations and background. Hence, the simulator is very useful for
investigating the effect of this variation. Fig.~\ref{ResidStitch}.
shows the simulated residuals of four different BiSON stations over
July 1st and part of July 2nd (the solar oscillations and background
have not been added). This trace represents the first attempt to
generate such a realistic data set and makes an interesting
comparison with previous simulations which assume constant noise.
The type and level of the noise sources have been matched as well as
possible to real data.

\section{Discussion}

A new BiSON simulator has been designed to more accurately mimic the
instrumental noise generated in the data. We will now use the
simulator to help us better understand the effect of instrumental
noise on the precise mode parameters extracted from power spectra
and on the detection of low power modes. In addition, our analysis
here shows that a time series produced from multiple BiSON stations
will have a varying noise profile suggesting that better statistics
may be achieved by applying a weight to each of the time series
points. This may be accomplished using a sine-wave fitting technique
- see New et al from the proceedings of this meeting.

\acknowledgements 
STF acknowledges the support of the Faculty of Arts, Computing,
Engineering and Science (ACES) at the University of Sheffield and
the support of the Science and Technology Facilities Council (STFC).
We also thank all those associated with BiSON which is funded by the
STFC.



\end{document}